\documentclass[11pt]{article}
\usepackage{moriond02,epsfig}

\def\be{\begin{equation}}
\def\ee{\end{equation}}
\def\bea{\begin{eqnarray}}
\def\eea{\end{eqnarray}}

\def\MET{\mbox{${\hbox{$E$\kern-0.6em\lower-.1ex\hbox{/}}}_T$}} 

\def\MP{\mbox{$M_P$}}
\def\mp{\mbox{$M_P$}\ }
\def\TH{\mbox{$T_H$}\ }
\def\mbh{\mbox{$M_{\rm BH}$}\ }     
\def\MBH{\mbox{$M_{\rm BH}$}}         
\def\MET{\mbox{${\hbox{$E$\kern-0.6em\lower-.1ex\hbox{/}}}_T$}} 
\def\met{\mbox{${\hbox{$E$\kern-0.6em\lower-.1ex\hbox{/}}}_T$}\ } 
\def\ifb{fb$^{-1}$}                     

\begin{document}
\vspace*{4cm}
\title{Black Holes at Future Colliders and Beyond~\footnote{Full version of this talk is available from {\tt http://hep.brown.edu/users/Greg/talks/Moriond02.pdf.}}}

\author{GREG LANDSBERG}

\address{Brown University, Department of Physics, 182 Hope St., Providence, RI 02912, USA\\E-mail: landsberg@hep.brown.edu}

\maketitle
\vspace*{-0.4in}
\abstracts{
One of the most dramatic consequences of low-scale ($\sim 1$~TeV) quantum gravity is
copious production of mini black holes at future accelerators and in ultra-high-energy cosmic ray interactions. Hawking radiation of these black holes is constrained mainly to our (3+1)-dimensional world and results in rich phenomenology. We discuss tests of Wien's law of Hawking radiation, which is a sensitive probe of the dimensionality of extra space, as well as an exciting possibility of finding new physics in the decays of black holes.
}
\vspace*{-0.1in}

\section{Introduction}

The possibility that the universe has more than three spatial dimensions has been discussed since it was first suggested by Riemann. String theory would have us believe that there could be up to $n=7$ additional dimensions, compactified at distances of the order of $10^{-32}$~m. In a new model,\cite{add} inspired by string theory, several of the compactified extra dimensions are suggested to be as large as 1~mm. These large extra dimensions are introduced to solve the hierarchy problem of the standard model (SM) by lowering the Planck scale ($M_{\rm Pl}$) to an energy range of a TeV. We refer to this {\it fundamental\/} Planck scale as $M_P$. 

As was pointed out in the past,\cite{adm,bf,ehm} an exciting consequence of TeV-scale quantum gravity is the possibility of production of black holes (BHs) at the accelerators. Recently, this phenomenon has been quantified for the case of TeV-scale particle collisions,\cite{dl,gt} resulting in a mesmerizing prediction that future colliders would produce mini black holes at an enormous rate, e.g. $~ 1$~Hz at the LHC for $M_P = 1$~TeV, thus becoming black-hole factories.

In this talk we review the phenomenology of the black hole production and decay in high-energy collisions and point out exciting ways of studying quantum gravity and searching for new physics using large samples of black holes accessible at future colliders.

\section{Assumptions}

Black holes are well understood general-relativistic
objects when their mass \mbh far exceeds the fundamental (higher
dimensional) Planck mass $\MP \sim$TeV. As \mbh approaches \MP,
the BHs become ``stringy'' and their properties complex. In what 
follows, we will ignore this obstacle 
and estimate the properties of light BHs by simple semiclassical 
arguments, strictly valid for $\MBH \gg \MP$. We expect that this 
will be an adequate approximation, since the important experimental
signatures rely on two simple qualitative properties: (i) the
absence of small couplings and (ii) the ``democratic" nature of BH 
decays, both of which may survive as average properties of the 
light descendants of BHs. We will focus on 
the production and sudden decay of Schwarzschild black holes.
The Schwarzschild radius $R_S$ of an $(4+n)$-dimensional black hole was derived in Ref.,\cite{mp} assuming that all $n$ extra dimensions are large ($\gg$ $R_S$).

As we expect unknown quantum gravity effects to play an increasingly 
important role for the BH mass approaching the fundamental Planck scale,
following the prescription of Ref.,\cite{dl} we do not consider BH 
masses below the Planck scale. It is expected that the BH 
production rapidly turns on, once the relevant energy threshold
$\sim\! M_P$ is crossed. (At lower energies, we expect BH production
to be exponentially suppressed due to the string excitations or 
other quantum effects.)  

\section{Black Hole Production and Decay}

Consider two partons with the center-of-mass energy $\sqrt{\hat s} =
\MBH$ moving in opposite directions. Semiclassical reasoning
suggests that if the impact parameter is less than the (higher
dimensional) Schwarzschild radius, a BH with the mass \mbh forms.
Therefore the total cross section can be estimated from
geometrical arguments and is of order $\pi R_S^2$.\footnote{Exponential 
suppression of the semiclassical formula, claimed in Ref.,\protect\cite{Voloshin} 
was not confirmed by subsequent studies~\protect\cite{de,nosuppression} and thus 
won't be discussed here.} 

Using the expression for the Schwarzschild radius,\cite{mp} we derive the following parton level BH production cross section:\cite{dl}
$$
    \sigma(\MBH) \approx \pi R_S^2 = \frac{1}{M_P^2}
    \left[
      \frac{\MBH}{\MP} 
      \left( 
        \frac{8\Gamma\left(\frac{n+3}{2}\right)}{n+2}
      \right)
    \right]^\frac{2}{n+1}.
$$

In order to obtain the production cross section in $pp$ collisions at the LHC, 
we use the parton luminosity approach:\cite{dl,gt,EHLQ}
$$
    \frac{d\sigma(pp \to \mbox{BH} + X)}{d\MBH} = 
    \frac{dL}{dM_{\rm BH}} \hat{\sigma}(ab \to \mbox{BH})
    \left|_{\hat{s}=M^2_{\rm BH}}\right.,
$$
where the parton luminosity $dL/d\MBH$ is defined as the sum over
all the types of initial partons:
$$
    \frac{dL}{dM_{\rm BH}} = \frac{2\MBH}{s} 
    \sum_{a,b} \int_{M^2_{\rm BH}/s}^1  
    \frac{dx_a}{x_a} f_a(x_a) f_b(\frac{M^2_{\rm BH}}{s x_a}),
$$
and $f_i(x_i)$ are the parton distribution functions (PDFs). We
used the MRSD$-'$~\cite{MRSD} PDFs with the $Q^2$ scale taken
to be equal to \MBH, which is within the allowed range of these
PDFs for up to the kinematic limit at the LHC. The dependence of 
the cross section on the choice of PDF is $\sim 10\%$.
The total production cross section for $\MBH > M_P$ at the LHC, 
obtained from the above equation, ranges between 15 nb and 1 pb for 
the Planck scale between 1 TeV and 5 TeV, and varies by $\approx 10\%$ 
for $n$ between 2 and 7.

Once produced, the TeV BHs quickly evaporate via Hawking 
radiation~\cite{Hawking} with a characteristic temperature
$$
    T_H = \MP
    \left(
      \frac{\MP}{\MBH}\frac{n+2}{8\Gamma\left(\frac{n+3}{2}\right)}
    \right)^\frac{1}{n+1}\frac{n+1}{4\sqrt{\pi}} = \frac{n+1}{4\pi R_S}
$$
of $\sim 100$~GeV~\cite{dl,gt}. The average multiplicity of particles 
produced in the process of BH evaporation is given by~\cite{dl,gt} and 
is of the order of half-a-dozen for typical BH masses accessible 
at the LHC. Since gravitational coupling is flavor-blind, a BH emits all the 
$\approx 120$ SM particle and antiparticle degrees of freedom with 
roughly equal probability. Accounting for color and spin, we expect 
$\approx 75\%$ of particles produced in BH decays to be quarks and gluons, 
$\approx 10\%$ charged leptons, $\approx 5\%$ neutrinos, and $\approx 5\%$ 
photons or $W/Z$ bosons, each carrying hundreds of GeV of energy. 
Similarly, if there exist new particles with the scale $\sim 100$~GeV, 
they would be produced in the decays of BHs with the probability 
similar to that for SM species. For example, a sufficiently light 
Higgs boson is expected to be emitted in BH decays with $\sim 1\%$ probability. This has exciting consequences for searches for new physics at the LHC and beyond, as the production cross section for any new particle via
this mechanism is (i) large, and (ii) depends only weakly on particle mass, 
in contrast with the exponentially suppressed direct production mechanism.

A relatively large fraction of prompt and energetic photons, electrons, 
and muons are expected in the high-multiplicity BH decays, which would 
make it possible to select pure samples of BH events, which are also 
easy to trigger on.\cite{dl,gt} At the same time, only small fraction 
of the particles produced in the BH decays are undetectable gravitons 
and neutrinos, so most of the BH mass is radiated in the form of visible 
energy.

\begin{figure}[tbp]
\begin{center}
\epsfxsize=5in
\epsffile{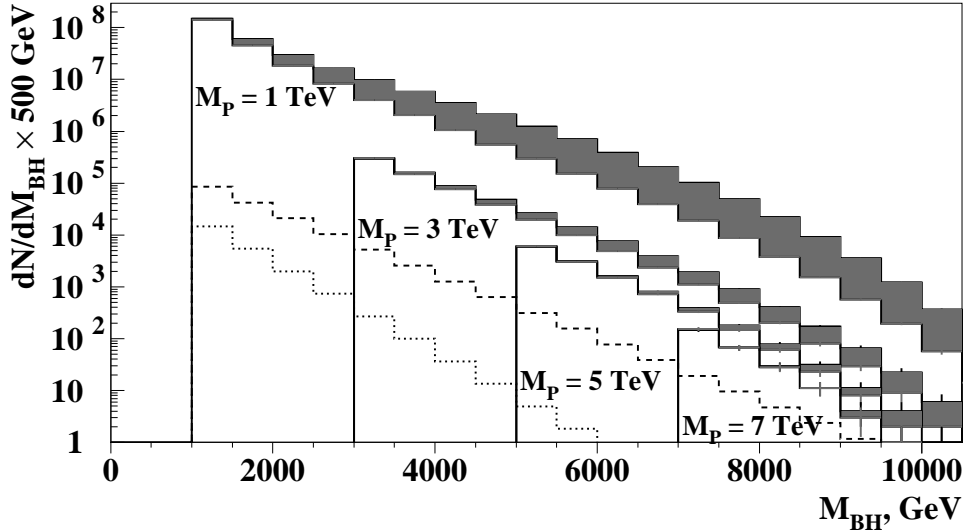}
\caption{Number of BHs produced at the LHC in the electron or photon decay 
channels, with 100~\protect\ifb of integrated luminosity, as a function of the BH 
mass. The shaded regions correspond to the variation in the number of events 
for $n$ between 2 and 7. The dashed line shows total SM background 
(from inclusive $Z(ee)$ and direct photon production). The dotted line 
corresponds  to the $Z(ee)+X$ background alone. From Ref.\protect\cite{dl}}
\label{nbh}
\end{center}
\end{figure}

In Fig.~\ref{nbh} we show the number of BH events tagged by a presence of an energetic electron or photon among the decay products in 100~fb$^{-1}$ of data collected at the LHC, together with the SM backgrounds, as a function of the BH mass.\cite{gl} It is clear that 
very clean and large samples of BHs can be produced up to Planck mass of
$\sim 5$ TeV. Note, that the BH discovery potential at the LHC is maximized in the
$e/\mu+X$ channels, where background is much smaller than that 
in the $\gamma+X$ channel (see Fig.~\ref{nbh}). The reach of a simple 
counting experiment extends up to $\MP \approx 9$ TeV ($n=2$--7),
where one would expect to see a handful of BH events with negligible 
background. 

\begin{figure}[htbp]
\begin{center}
\epsfxsize=5in
\epsffile{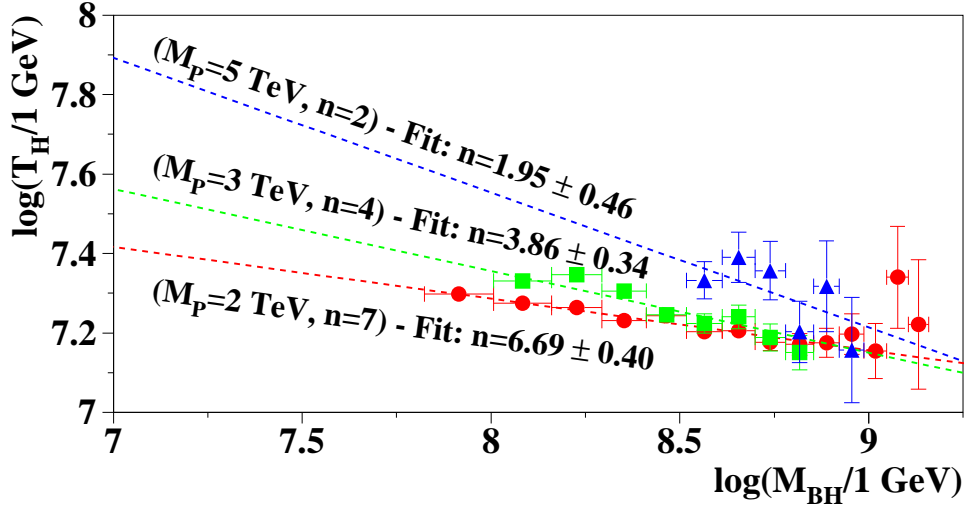}
\caption{Determination of the
dimensionality of  space via Wien's displacement law at the LHC with
100~\protect\ifb\ of data. From Ref.\protect\cite{dl}}
\label{Wien}
\end{center}
\end{figure}

A sensitive test of properties of Hawking radiation can be obtained by measuring the relationship between the mass of the BH (reconstructed from the total energy of all the decay products) and its Hawking temperature (measured from the energy spectrum of the decay pro\-ducts). We can use the measured \mbh vs. \TH\ dependence to determine both the fundamental Planck scale \mp and the dimensionality of space $n$. This is a multidimensional equivalent of the Wien's law. Note that determination of $n$ can be done in largely model-independent way via taking a logarithm of both parts of the expression for Hawking temperature: $\log(T_H) = -\,\frac{1}{n+1}\log(\MBH) + \mbox{const}$, 
where the constant does not depend on the BH mass, but only
on \mp and on detailed properties of the bulk space, such as shape of
extra dimensions.\cite{gl} Therefore, the slope of a straight-line fit to
the $\log(\TH)$ vs. $\log(\MBH)$ data offers a direct way of determining
the dimensionality of space. The reach of this method at the LHC is illustrated in Fig.~\ref{Wien} and discussed in detail in Ref.\cite{dl} Note that the determination of the dimensionality of space-time by this method is fundamentally different from other ways of determining $n$, e.g.
by studying a monojet signature or a virtual graviton exchange processes,
also predicted by theories with large extra dimensions. 

\section{Discovering New Physics in the Decays of Black Holes}

As an example,\cite{gl} we study the discovery potential of the BH sample collected at the LHC for a SM-like Higgs boson with the mass of 130 GeV, still allowed in low-scale 
supersymmetry models, but very hard to establish at the Fermilab Tevatron.\cite{SUSYHiggs} We consider decay of the Higgs boson into pair of jets, dominated by the $b\bar b$ final state (57\%), with an additional 10\% contribution from the $c\bar c$, $gg$, and 
hadronic $\tau\tau$ final states.

We model the production and decay of the BH with the TRUENOIR Monte Carlo 
generator,\cite{Snowmass} which implements a heuristic algorithm to describe a 
spontaneous decay of a BH. The generator is interfaced with the PYTHIA Monte 
Carlo program~\cite{PYTHIA} to account for the effects of initial and final state radiation, particle decay, and fragmentation. We used a 1\% probability 
to emit the Higgs particle in the BH decay. We reconstruct final state particles within the acceptance of a typical LHC detector and smear their energies with typical resolutions. 

\begin{figure}[tbp]
\begin{center}
\epsfxsize=5in
\epsffile{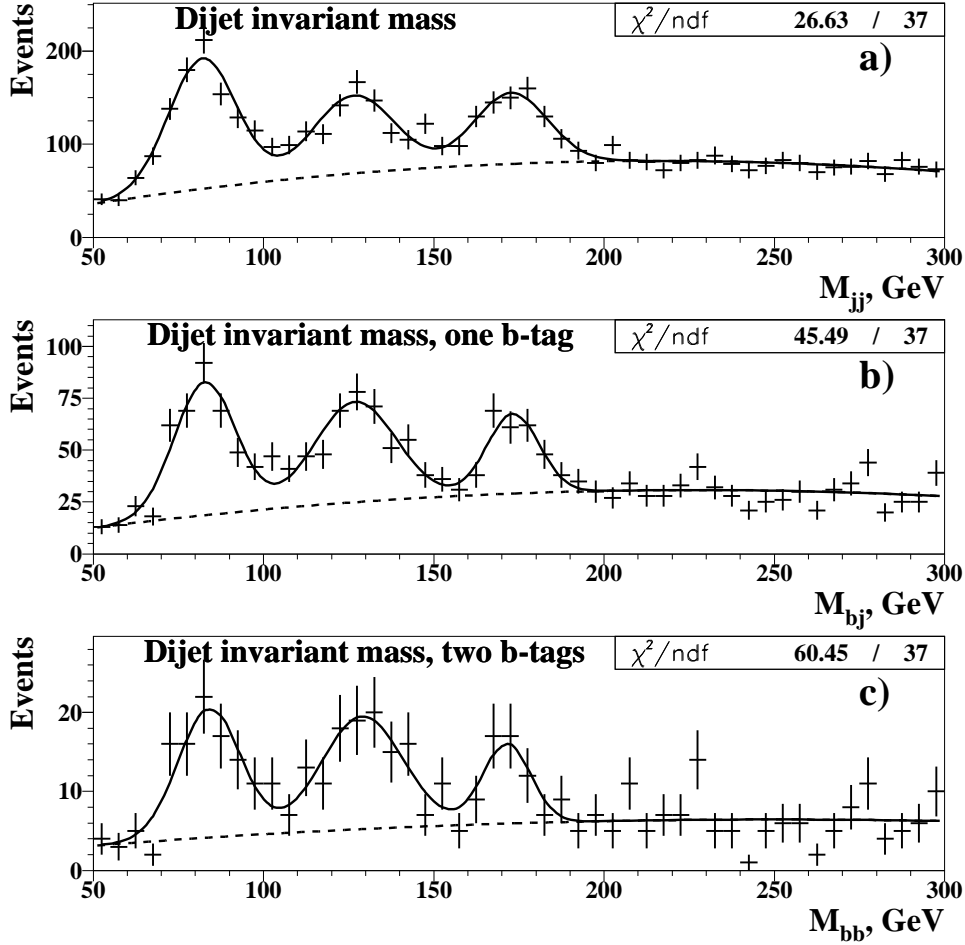}
\caption{Dijet invariant mass observed in the BH decays with a prompt 
lepton or photon tag in $\approx$3~pb$^{-1}$ of the LHC data, for 
$M_P = 1$~TeV and $n=3$: (a) all jet combinations; (b) jet combinations 
with at least one of the jets tagged as a $b$-jet; (c) jet 
combinations with both jets tagged as $b$-jets. The solid line is a 
fit to a sum of three Gaussians and a polynomial background, shown 
with the dashed line. The three peaks correspond to the $W/Z$ bosons, 
the Higgs boson, and the top quark (see text). The $\chi^2$ per d.o.f.
is shown to demonstrate the quality of the fit. Note, that as the 
maximum likelihood fit was used for all cases, the $\chi^2$ in (c) is 
not an appropriate measure of the fit quality due to low statistics. 
Using the Poisson statistics, the probability of the fit (c) is 8\%. 
From Ref.\protect\cite{gl}}
\label{bhhiggs}
\end{center}
\end{figure}

The simplest way to look for Higgs in the BH decays is to use the 
dijet invariant mass spectrum for all possible combinations of jets found among the final state products. This spectrum is shown in Fig.~\ref{bhhiggs} for $M_P = 1$~TeV and $n=3$. The three panes correspond 
to all jet combinations (with the average of approximately four jet 
combinations per event), combinations with at least one $b$-tagged jet, 
and combinations with both jets $b$-tagged. The most prominent feature 
in all three plots is the presence of three peaks with the masses around 
80, 130, and 175 GeV. The first peak is due to the hadronic decays of the 
$W$ and $Z$ bosons produced in the BH decay either directly or in the decays 
of the top and Higgs particles. (The resolution of a typical LHC detector 
does not allow to resolve $W$ and $Z$ in the dijet mode.) The second peak 
is due to the $h \to jj$ decays, and the third peak is due to the $t \to 
Wb \to jjb$ decays, where the top quark is highly boosted. In this case, 
one of the jets from the $W$ decay sometimes overlaps with the prompt 
$b$-jet from the top quark decay, and thus the two are reconstructed as 
a single jet; when combined with the second jet from the $W$ decay, this 
gives a dijet invariant mass peak at the top quark mass. The data set
shown in Fig.~\ref{bhhiggs} consists of 50K BH events, which, given the 15~nb 
production cross section for $M_P = 1$~TeV, $n = 3$, corresponds to an integrated luminosity of 3 pb$^{-1}$, or less than an hour of the LHC operation at the nominal 
luminosity. The significance of the Higgs signal shown in Fig.~\ref{bhhiggs}a 
is 6.7$\sigma$, even when no $b$-tagging is involved.

With this method, the $5\sigma$ discovery of a 130 GeV Higgs boson may be possible with 
${\cal L} \approx 2$~pb$^{-1}$ (first day), 100~pb$^{-1}$ (first week), 1~fb$^{-1}$ (first month), 10~fb$^{-1}$ (first year), and 100~fb$^{-1}$ (full LHC sample) for the fundamental Planck scale of 1, 2, 3, 4, and 5~TeV, respectively, even with incomplete and poorly calibrated LHC detectors. The integrated luminosity required is significantly lower than that for direct Higgs boson production, if the Planck scale is below 4 TeV. While the studies were done for a particular Higgs boson mass, the dependence on the mass is small. Moreover, this approach is applicable to searches for other new particles with the masses $\sim 100$~GeV, e.g. low-scale supersymmetry. Light slepton or top squark searches via 
this technique may be particularly fruitful. Very similar conclusions 
apply not only to BHs, but to other possible intermediate quantum states, 
such as string balls,\cite{de} which have similar production cross section 
and decay modes as BHs. In this case, the relevant mass scale is not the 
Planck scale, but the string scale, which determines the evaporation 
(Haggedorn) temperature.\cite{de}

\section{Black Holes in Cosmic Rays}

\begin{figure}[thbp]
\begin{center}
\epsfxsize=5in
\epsffile{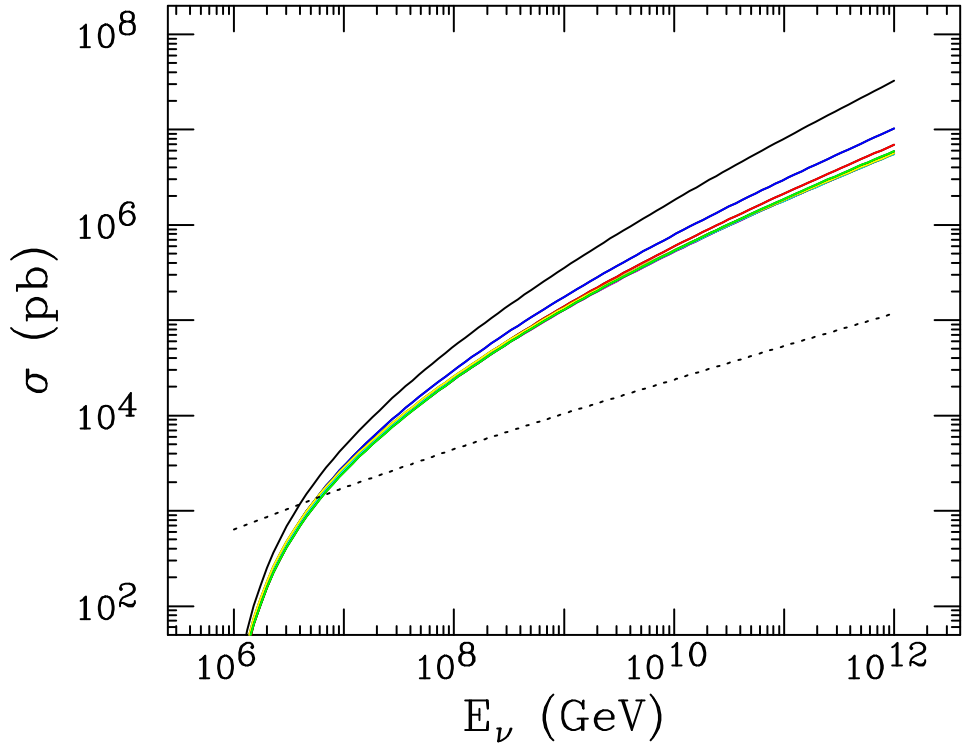}
\caption{Cross sections $\sigma ( \nu N \to {\rm BH})$ for $M_P =
M_{\rm BH}^{\rm min} = 1$~TeV and $n=1, \ldots, 7$. (The last four
curves are virtually indistinguishable.)  The dotted curve is for the
SM process $\nu N \to \ell X$. From Ref.\protect\cite{Feng}}
\label{bhcr}
\end{center}
\end{figure}

It was suggested recently that mini black hole production can be observed also in the interactions of Earth-skimming ultra-high energy neutrinos.\cite{Feng} For neutrino energies above $\sim 10^7$ GeV, the BH production cross section by neutrinos would exceed their SM interaction rate (see Fig.~\ref{bhcr}). Several ways of detecting such interactions have been proposed,\cite{Feng,CR} including large-scale ground-based arrays, space probes, and neutrino telescopes. If the Planck scale is sufficiently low (1-3 TeV) a handful of BH events could be observed by, e.g. Pierre Auger observatory even before the LHC turns on. However, it would not be easy to establish the BH nature of the event excess in cosmic ray interactions unambiguously due to limited particle identification capabilities of these detectors.

\section{Conclusions}

To conclude, black hole production at the LHC may be one of the early
signatures of TeV-scale quantum gravity. It has three advantages:
\begin{enumerate}
\item Large Cross Section. Because no small dimensionless coupling constants, analogous to $\alpha$, suppress the production of BHs. This leads to enormous rates.

\item Hard, Prompt, Charged Leptons and Photons. Because thermal decays are flavor-blind. This signature has practically vanishing SM background.

\item Little Missing Energy. This facilitates the determination of the mass and the temperature of the black hole, and may lead to a test of Hawking's radiation.
\end{enumerate}
Large samples of black holes accessible by the LHC would allow for precision 
determination of the parameters of the bulk space and may result in the discovery of new particles in the BH evaporation. Small samples of BH events may be observed in ultra-high-energy cosmic ray experiments, even before the LHC turns on. 

If the models with large extra dimensions are realized in nature, the production of BH in the lab is just years away. That would mark an exciting transition for astroparticle physics, truly unifying it with cosmology.

\section{Acknowledgments}

I would like to thanks the Moriond organizers for the invitation and for an exciting conference. I am especially grateful to my coauthor on the first black hole paper,\cite{dl} Savas Dimopoulos, for a number of stimulating discussions and support. This work has been supported partially by the U.S.~Department of Energy under Grant No. DE-FG02-91ER40688 
and by the Alfred P. Sloan Foundation.

\end{document}